\documentclass[aoas,MSNbibl,nameyear,dvips]{arximspdf}
\usepackage{graphicx}

\doi{10.1214/12-AOAS544} %kopijuoti is PTS
\volume{6}
\issue{3}
\pubyear{2012}
\firstpage{853}
\lastpage{869}



\begin{document}
\begin{frontmatter}

\title{Matching markers and unlabeled configurations in~protein~gels}
\runtitle{Matching partially labeled configurations}

\begin{aug}
\author{\fnms{Kanti V.}~\snm{Mardia}\ead[label=e2]{k.v.mardia@leeds.ac.uk}},
\author{\fnms{Emma M.}~\snm{Petty}\ead[label=e3]{emma.m.petty@googlemail.com}\thanksref{t1}}
\and
\author{\fnms{Charles C.}~\snm{Taylor}\corref{}\ead[label=e1]{c.c.taylor@leeds.ac.uk}}
\thankstext{t1}{Supported by a \textsc{case} studentship funded by the Engineering
and Physical Science Research Council and Central Science Laboratories, York, UK.}
\runauthor{K. V. Mardia, E. M. Petty and C. C. Taylor}
\affiliation{University of Leeds}
\address{Department of Statistics\\
University of Leeds\\
Leeds LS2 9JT\\
United Kingdom\\
\printead{e2}\\
\phantom{E-mail: }\printead*{e3}\\
\phantom{E-mail: }\printead*{e1}} %adresu isvedimo komanda gale!
\end{aug}

% HISTORY:
\received{\smonth{8} \syear{2010}}
\revised{\smonth{12} \syear{2011}}

% ABSTRACT
%
\begin{abstract}%\label{abstract}
Unlabeled shape analysis is a rapidly emerging and challenging area of
statistics. This has been driven by various novel applications in
bioinformatics. We consider here the situation where two configurations
are matched under various constraints, namely, the configurations have
a subset of manually located ``markers'' with high probability of
matching each other while a larger subset consists of unlabeled points.
We consider a plausible model and give an implementation using the EM
algorithm. The work is motivated by a real experiment of gels for renal
cancer and our approach allows for the possibility of missing and
misallocated markers. The methodology is successfully used to
automatically locate and remove a grossly misallocated marker within
the given data set.
\end{abstract}

% KEYWORDS
%
\begin{keyword}
\kwd{Electrophoresis}
\kwd{shape}
\kwd{Western Blots}.
\end{keyword}

\end{frontmatter}
%

%s1 ###
\section{Introduction}\label{intro}
%s1.1 ###
\subsection{Western Blots}\label{sec1.1}
\begin{figure}

\includegraphics{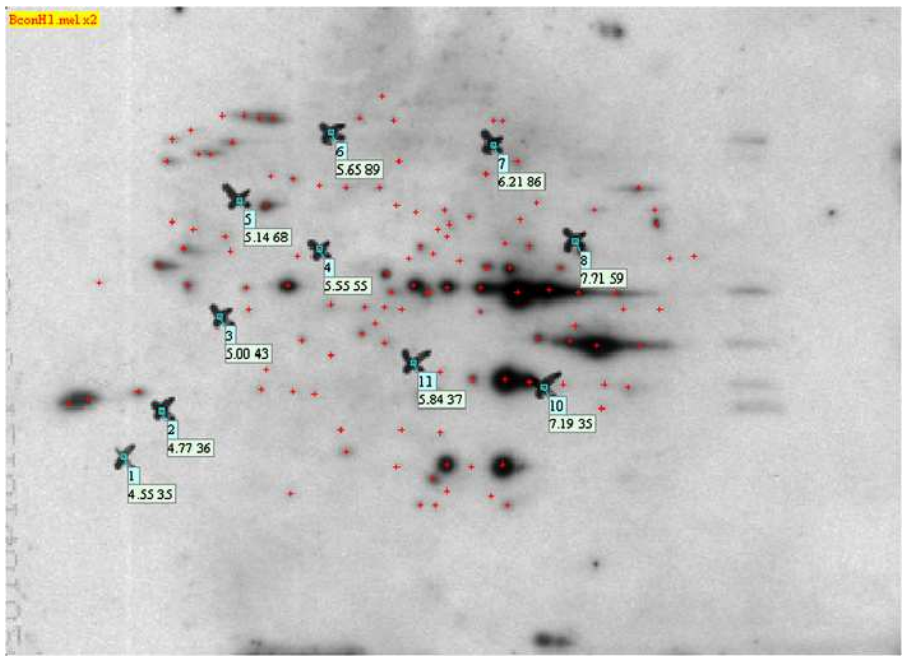}

\caption{Western Blot image with red crosses depicting the subject-treatment
specific nonmarkers. The larger black crosses indicate the labeled
markers, with their acidity and mass measurements (not
spatial coordinates) highlighted beneath.}\label{fig1:Image}
\end{figure}

Our motivating application concerns gel techniques used to identify
proteins present in human tissue. First, two-dimensional
electrophoresis (2-DE) is used to separate all the proteins extracted
from a~cell. The 2-DE gel is then probed with serum which contains
antibodies that will bind to specific proteins. The image of a Western
Blot will contain only the location (and intensity)
of proteins that have a bound antibody. We can think of Western Blots
as containing only a subset of the proteins that are displayed on 2-DE
images. The extra step necessary to create a Western Blot allows a
further level of variability within the final image. The
reproducibility of Western Blots is therefore even more challenging
than that of 2-DE images.
To help align Western Blots, suitable marker proteins are
experimentally determined and are generally expected to be present in
all blots under investigation. A stain is applied to each blot which
will highlight all proteins present, therefore enabling an expert to
manually locate the suitable markers. Figure~\ref{fig1:Image} shows an
annotated Western Blot
image which shows the markers (with the acidity and mass measurements
associated with these points) and further points
detected by an image analyzer.
The markers are used to align the blots by minimizing a sum of
squared euclidean distances (usually not the acidity and mass measurements).
In some cases, fine adjustments to alignments are made using various
heuristic techniques. See, for example, Forgber et al. (\citeyear{Foretal09}) and
Zvelebil and Baum [(\citeyear{ZveBau}), pages~613--620] for more details.
%
%f1 ###

Considering the large scope for variation between images and the often
vast number of proteins located in a comparatively small area, visual
examination to analyze or compare images, although often informative,
can be extremely difficult and conclusions unreliable. Visual
comparison can also be extremely repetitive and laborious for the
expert making the comparisons. Statistical and computational analysis
is essential to the \textit{result accuracy} and reduction of expert
manual labor. The main aim is to locate a biomarker whose mere presence
can be used to measure the progress of disease or treatment effects. In
the case of the gel data, a~point becomes a biomarker if it is found to
have this property. The intensity of a
biomarker, indicated by the intensity of the mark on the image, can
also provide information about
the disease progression or treatment effect, but this is beyond the
scope of this paper.

%s1.2 ###
\subsection{Unlabeled configuration matching}\label{sec1.2}
In the more general setting, the problem is to match two sets---usually of unequal size---of points, in which
the correspondence (matching) of the points is unknown. The solution
will include the transformation
required to align the sets, a list of correspondences which map (some
of) the points, and will penalize solutions with many unmatched points,
allowing for a trade-off in the goodness of fit in
the aligned points.

Approaches to closely related problems include the RANSAC algorithm
[Fischler and Bolles (\citeyear{FisBol81})], nonrigid point matching using thin-plate splines
[Chui and Rangarajan (\citeyear{ChuRan03})], a correlation-based approach using kernels
[Tsin and Kanade (\citeyear{TsiKan04}), Chen (\citeyear{Che11})], nonaffine matching of distributions
[Glaunes, Trouv{\'e} and Younes (\citeyear{GlaTroYou04})] and the Iterative Closest Point Algorithm
[Besl and McKay (\citeyear{BesMcK92})] for the registration of various representations of
shapes. All of these methods avoid making distributional assumptions,
with a consequence that probabilistic statements are then difficult to
make. By contrast, Czogiel, Dryden and Brignell (\citeyear{CzoDryBri}),
Dryden, Hirst and Melville (\citeyear{DryHirMel07}), Kent, Mardia and Taylor (\citeyear{KenMarTay10N1}),
Taylor, Mardia and Kent (\citeyear{TayMarKen03}) and Green and Mardia (\citeyear{GreMar06})
use statistical models to obtain solutions.
These latter papers all use examples drawn from protein bioinformatics;
a~review is given by
Green et al. (\citeyear{Greetal10}).

In this paper we address a more specific problem in which each
configuration contains
a subset of points (``markers'') whose labels correspond with high
probability, with the
remaining points having arbitrary labels (nonmarkers) as before.
Suppose we have two configurations of observed landmarks in
$d$ dimensions: markers given by $x_j$, $j=1, \ldots, K$ and $\mu_i$, $i=1, \ldots, K$,
and nonmarkers
$\mu_i$, $i = K+1 , \ldots, K+m$ and $x_j$, $j = K+1 , \ldots, K+n$.
These are represented as matrices $x ((K+n) \times d)$ and $\mu ((K+m) \times d)$ in which $K$ is usually
smaller than $m$ and $n$. In our model, the markers (the spatial
coordinates of the large black
crosses in Figure~\ref{fig1:Image}) $\mu_i$ and $x_i$ for $i=1,\ldots
,K$ have been identified by
an expert to correspond to the same proteins (referred to as a
``points'' hereafter). However,
these are labeled with some uncertainty, so true correspondence is
likely but not guaranteed.
So it is possible, for example, that markers in $\mu$ could correspond
to nonmarkers
in $x$, or have no correspondence at all. For $\mu_i$ and $x_j$ with
$i,j >K$, (the spatial
coordinates of the red crosses in Figure~\ref{fig1:Image}) we have no
prior information about correspondence probabilities.

%s1.3 ###
\subsection{Statistical model}\label{sec1.3}
\begin{figure}

\includegraphics{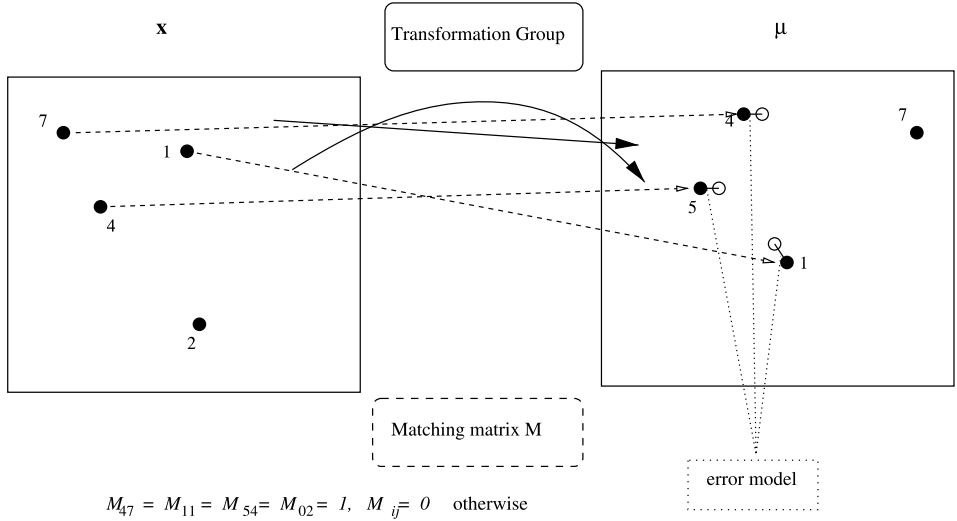}

\caption{Illustration of the main ingredients of a statistical model. The labels
of the two
configurations of points ($x$ and $\mu$) can be considered as
arbitrary. Some of the $x$ points
are aligned to some of the $\mu$ points by a transformation (e.g., translation, rotation)
which belongs to a specified group. An $0/1$ matrix $M$ indicates which
points match,
with unmatched points in $x$ (point 2 in the illustration) assigned to
label ``0,'' and a
specific error model assumed for the magnitude of the residual after alignment.}\label{fig-gen}
\end{figure}

A statistical model in the general setting involves three main
components (see Figure~\ref{fig-gen}):
\begin{longlist}[(a)]
\item[(a)] A group $\mathcal{G}$, say, on
$\mathbb{R}^d$ representing the permitted transformations~($g$) on (a~subset of the landmarks of) $\mu$ to bring it close to (a subset of
the landmarks of) $x, g \in\mathcal{G}$.

\item[(b)] A matching matrix $M$, say, identifies which
elements of $x$ correspond to which elements of $\mu$ for the markers
as well as unlabeled points.

\item[(c)] An error model indicating how close the elements of $x$
and $\mu$ will be, after the correct transformation and labeling
are used.
\end{longlist}
%
%f2 ###

In Section~\ref{statmodel} we introduce our statistical model and emphasize the group
of affine transformations belonging to $\mathcal{G}$ which is relevant
to our example. The appropriate matching matrix $M$
is estimated under various scenarios, including the use of a matrix $Q$
of prior probabilities,
which is introduced to reflect the existence of the markers (labeled
points)---an integral part of the specific problem. In Section~\ref{sec3} we outline
likelihood based inference for~$M$, and describe an EM algorithm.
% Also, we introduce priors for $\mu$ under different matching schemes.
In Section~\ref{sec4} we adapt the prior matrix~$Q$ when either a marker is
missing or a marker is
wrongly identified. Two real examples are studied in Section~\ref{sec5} related
to renal cancer. In the first example, one marker is grossly
misallocated and in the second example, some markers are missing. This
procedure has great potential to automate preprocessing of the gels. We
conclude with a discussion.\looseness=1

%%%%%%%%%%%%%%%%%%%%%%%%%%%%%%%%%%%%%%%%%

%s2 ###
\section{Statistical models}\label{statmodel}

%s2.1 ###
\subsection{Transformations}\label{sec2.1}

Although the statistical model we later introduce can apply to various
types of transformations, we focus on an affine transformation of the form
$g(\mu)=\mu A' + B'$,
where $A$ is a nonsingular $d \times d$ matrix and the $d \times1$
vector, $b$, is present in every column of the $d \times(K+m)$ matrix~$B$.

%s2.2 ###
\subsection{Matching matrix}\label{matching}
To estimate the parameters of an appropriate transformation of $\mu$,
we can introduce a correspondence system that will indicate whether a
point in $\mu$ is associated with
a point in $x$, that is, whether two points \textit{match} across configurations.
We can record the correspondence information in a $(K+m+1) \times
(K+n)$ matching matrix, $M$, where
\[
M_{ij} =
\cases{
1, &\quad $\mbox{for } i=0 \mbox{ if } x_j \mbox{ does not have a
matching point in } \mu,$\vspace*{2pt}\cr
1, &\quad $\mbox{for } i=1,\ldots,K+m \mbox{ if } x_j \mbox{ matches }
\mu_i,$\vspace*{2pt}\cr
0, &\quad $\mbox{otherwise}$}
\]
for $j=1,\ldots,K+n$. Note that, for simplicity of notation, we use
$M_{0j}\equiv M_{K+m+1,j}$,
and similarly for other matrices. If $M_{0j}=1$, then $x_j$ does not
have a matching point in $\mu$ and we say that $x_j$ is unmatched.

We consider one-to-one or many-to-one matches between points in $x$ and
points in $\mu$. We refer to these as \textit{hard} and \textit{soft}
matches,
respectively. Soft matching can be useful in our application
since a single protein can produce multiple spots on an image [Banks
et al. (\citeyear{Banetal00})].

\textit{Hard matches}:
The matching matrix, $M$, has the following constraints for the hard model:
%
%e1 ###
\begin{equation}
\sum_{i=0}^{K+m} M_{ij} = 1 \qquad\mbox{for }  j=1,\ldots,K+n
\label{eq2:MConstraint}
\end{equation}
and
%
%e2 ###
\begin{equation}
\sum_{j=1}^{K+n} M_{ij} \le1\qquad \mbox{for }  i=1,\ldots,K+m.
\label{eq2:HardMConstraint}
\end{equation}
So for $i_1 \neq0$, if $M_{i_1 j_1}=1$, then $M_{i_1 j_2} = M_{i_2
j_1}=0$ for all $i_1 \neq i_2$ and $j_1 \neq j_2$.
Note that there are no constraints on row $K+m+1$ in $M$ since each of
the $K+n$ points in $x$ is free to remain unmatched. Figure \ref
{fig-gen} illustrates the case of hard matches in which the point $x_2$
is unmatched, so $M_{02}=1$.

\textit{Soft matches}:
For the soft model, the only constraint is stated in (\ref
{eq2:MConstraint}). That is, if $M_{i_1 j_1}=1$, then $M_{i_2 j_1}=0$
for all $i_1 \neq i_2$, but $M_{i_1 j_2} \in\{ 0,1 \}$ for
$j_1 \neq j_2$.
When assigning either hard or soft matches, (\ref{eq2:MConstraint})
constrains a point in $x$ to be matched to a single point in $\mu$ or,
alternatively, to remain unmatched.

%s2.3 ###
\subsection{Error distribution}\label{sec2.3}
Assuming the transformation parameters,\break $A$ and~$b$, are known, we can
apply a distribution to $x_j$ given the match $M_{ij}=1$. Given the
transformation, we treat the elements of $x$ as conditionally
independent with the following densities for $j=1,\ldots,K+n$:
%
%e3 ###
\begin{equation}\label{eq2:x|M}
p(x_j | M_{ij}=1) =
\cases{
\displaystyle\frac{1}{(2\pi\sigma^2)^{d/2}} \exp\biggl\{ -\frac{\| x_j - A\mu_i-b\|
^2}{2\sigma^2} \biggr\},\vspace*{2pt}\cr
\hspace*{33pt}\quad\mbox{for } i = 1,\ldots,K+m,\vspace*{2pt}\cr
\displaystyle\frac{1}{| \Omega|},  \qquad\mbox{for } i=0,}
\end{equation}
where $\Omega$ is some region in $\mathbb{R}^d$ containing all points
in $x$.

To allow for the possibility of soft matching, we consider points in
$x$ to be independent. As we have $K$ markers in each image, we have
prior information about the matching across images. Next we introduce
notation to deal with prior matching probabilities.

%s2.4 ###
\subsection{Prior matching matrix probabilities}\label{sec2.4}
Let $Q$ be a $(K+m+1) \times(K+n)$ matrix with elements $q_{ij} =
p(M_{ij}=1)$. That is, for $j=1,\ldots,K+n$, $q_{ij}$ is the prior
probability that $\mu_i$ is matched to $x_j$ for $i = 1,\ldots, K+m$
and the prior probability that $x_j$ is unmatched for $i=0$. Again, for
simplicity of notation, we use $q_{0j}$ in place of $q_{K+m+1,j}$.
Note that $\sum_{i=0}^{K+m} q_{ij} = 1 \mbox{ for } j=1,\ldots,K+n$.
We have prior knowledge that corresponding markers, $\mu_j$ and~$x_j$
for $j=1, \ldots, K$, \textit{should} match.
We propose a structure to determine the~$q_{ij}$, which accounts for
the possibility of error when allocating markers within a~warped image
and does not force corresponding markers to match.
In what follows, it will be helpful to note that the matrix $Q$ can be
partitioned into
submatrices of size (rows $\times$ columns) as follows:
\begin{eqnarray*}
&&Q \bigl((1+K+m)\times(K+n)\bigr)\\
&&\qquad=
\pmatrix{
Q^{(0)}(1 \times K) & \mid& \vspace*{2pt}\cr
--------- & \mid\vspace*{2pt}\cr
&\mid& Q^{(2)}\bigl((1+K+m) \times n\bigr)\vspace*{2pt}\cr
Q^{(1)}\bigl((K+m) \times K\bigr)&\mid\vspace*{2pt}\cr
& \mid& }.
\end{eqnarray*}

\textit{Markers in} $x$:
We know that $\mu_j$ are the coordinates for marker $j$ in $\mu$,
$j=1,\ldots,K$. Let $\gamma_j$ be the index of the true marker $j$ in
$\mu$. If $\gamma_j = j$, then the marker $j$ has been correctly identified.
We set the prior probability of a point $\mu_i$ being the true marker
$j$, $q_{ij}$, to be a function of the distance between $\mu_i$ and $\mu
_j$ so that $Q^{(1)}$ has elements
%
%e4 ###
\begin{equation}
q_{ij} = p(\gamma_j = i) = f(d_{ij}) \qquad \mbox{for }
i=1,\ldots,K+m,  j=1,\ldots, K,
\label{eq2:qAdaptM}
\end{equation}
where $d_{ij}$ is the Euclidean distance between $\mu_i$ and $\mu_j$
and choices for $f$ are discussed later.\vadjust{\goodbreak}

Next we consider the possibility that a marker within $x$ does not have
a~corresponding point in $\mu$. Recall that $x_j$ are the coordinates
for marker $j$ in~$x$, $j=1,\ldots,K$. To allow for the possibility
that $x_j$ remains unmatched, we set the prior probability of
$M_{0j}=1$ to be uniform so that $Q^{(0)}$ has elements\looseness=1
%
%e5 ###
\begin{equation}
q_{0j} = p(\gamma_j = 0) = \frac{1}{| \Omega|}\qquad \mbox
{for } j=1,\ldots, K,
\label{eq2:qAdapt0}
\end{equation}\looseness=0
where $\Omega$ is given as in (\ref{eq2:x|M}).

\textit{Nonmarkers in} $x$:
To allow for matching of the nonmarker points, we can set the elements
of $Q^{(2)}$ as
%
%e6 ###
\begin{equation}
q_{ij} = \frac{1}{K+m+1},\qquad  i=0,\ldots,K+m,  j=K+1,\ldots,K+n.
\label{eq2:qAdaptNM}
\end{equation}
So the prior matching probability of a nonmarker $x_j$ is uniform.

As an example, we suppose that in Figure~\ref{fig-gen} only point 1 has
been identified as a marker in
both $x$ and $\mu$, then we might have $q_{01}=0.01$ ($=1/|\Omega|$,
say), $q_{11}=0.89$,
$q_{41}=0.01$, $q_{51}=0.09$, $q_{71}=0.00$ (based on the interpoint
distances within $\mu$) and
$q_{ij}=1/8$ for the other points shown (taking $m=6$ in this example).

For ease of reference, the ingredients of the statistical model,
together with
possible variations, are listed in Table~\ref{tab1}.

%t1 ###
\begin{table}
\caption{Main ingredients of the statistical model used
for matching of partially~labeled~configurations of points.
Section numbers [e.g., (\protect\ref{sec3.1})] are
used~to~sign-post further details or discussion}\label{tab1}
\begin{tabular*}{\textwidth}{@{\extracolsep{\fill}}lcc@{}}
\hline
\multicolumn{1}{@{}l}{\textbf{Component of model}} & \textbf{Variants} & \textbf{Examples}\\
\hline
\multicolumn{1}{@{}l}{Configurations $x$ and $\mu$} & Unlabeled (Section
\ref{sec1.2})\\
 &Partially labeled & Markers (Section~\ref{sec1.1})\\
\multicolumn{1}{@{}l}{Transformation group} & Rigid-body (Section~\ref{sec2.1})\\
& Affine (Section~\ref{sec3.1})&\\
&Nonlinear (Section~\ref{sec6})&\\
\multicolumn{1}{@{}l}{Matching matrix, $M$} & Hard (Section~\ref{sec6}) &
$\!\!\!\!\!\!$One-to-one\\
&Soft & $\!\!\!\!\!\!$Many-to-one (Section~\ref{sec6})\\
&&$\!\!\!\!\!\!$Many-to-many (Section~\ref{sec6})\\
\multicolumn{1}{@{}l}{Prior matrix, $Q$,}&&\\
\multicolumn{1}{@{}l}{with $Q_{ij}=P(M_{ij}=1)$}&&\\
\multicolumn{1}{@{}l}{which depends on}&&\\
\quad -- markers (Section~\ref{sec4}) & Function of distance (Section~\ref{sec3.3.1})&\\
\quad -- nonmarkers&&\\
\multicolumn{1}{@{}l}{Error distribution} & Isotropic (Section~\ref{sec2.3})& \\
& Nonlinear (Section~\ref{sec6})&\\
\hline
\end{tabular*}
\end{table}

%%%%%%%%%%%%%%%%%%%%%%%%%%%%%%%%%%%%%%%%%%%%%%
%s3 ###
\section{EM algorithms and inference}\label{sec3}

%s3.1 ###
\subsection{EM algorithm}\label{sec3.1}

We use an EM algorithm [McLachlan and Krishnan (\citeyear{McLKri08})] to estimate the
transformation parameters,
$A$ and $b$, that will superimpose~$\mu$ onto $x$. Throughout this
section we assume that $\sigma^2$ has been assigned (see Section~\ref{sec3.3.3}).
In the E-step we calculate the posterior probability that $\mu_i$
matches~$x_j$, that is, the posterior probability that $M_{ij}=1$.
In the M-step the posterior probabilities are input into the expected
likelihood of observing
$M$, given the data, $x$. This enables us to estimate the
transformation parameters, $A$ and $b$.

\textit{E-step}:
We calculate the posterior probability of $\mu_i$ matching $x_j$, given
the data, using Bayes' theorem:
%
%e7 ###
\begin{equation}
p (M_{ij}=1 | x_j) = \frac{p(x_j | M_{ij}=1) p(M_{ij}=1)}{p(x_j)},
\label{eq2:Estep}
\end{equation}
where $p(x_j | M_{ij}=1)$ is calculated using (\ref{eq2:x|M}), and
$q_{ij}=p(M_{ij}=1)$ is calculated using (\ref{eq2:qAdaptM})--(\ref
{eq2:qAdaptNM}). The denominator of (\ref{eq2:Estep}) is given by
$\sum_{i=0}^{K+m} p(x_j | M_{ij}=1) \times p(M_{ij}=1)$.

\textit{M-step}: Starting from the multinomial form [McLachlan and Krishnan
(\citeyear{McLKri08}), page~15]
\[
l(M|x) = \sum_{i=0}^{K+m} \sum_{j=1}^{K+n} M_{ij} \log p(x_j),
\]
%
% | M_{ij}=1),$$
we
substitute $p_{ji}$ for $M_{ij}$ and $q_{ij}p(x_j | M_{ij}=1)$ for $p
(x_j)$% |M_{ij}=1)$
to obtain the expected
log-likelihood of the matching matrix, $M$, given the data, $x$:
%
%e8 ###
\begin{equation}
\mathsf{E}[l(M|x)] = \sum_{i=0}^{K+m} \sum_{j=1}^{K+n} p_{ji} [
\log q_{ij} + \log p(x_j | M_{ij}=1) ].
\label{eq2:ExLogLike2}
\end{equation}
Here, we suppress the dependence on the parameters $A$ and $b$.

Both the prior probabilities stored in $Q$ and the conditional
distribution of $x_j$ being unmatched are independent of $A$ and $b$,
so, using (\ref{eq2:ExLogLike2}), we estimate the transformation
parameters that maximize
\begin{eqnarray*}
&&\sum_{i=1}^{K+m} \sum_{j=1}^{K+n} p_{ji} \log p(x_j | M_{ij}=1) \\
&&\qquad= \sum
_{i=1}^{K+m} \sum_{j=1}^{K+n} p_{ji} \biggl[ -\frac{\| x_j - A\mu_i-b\|
^2}{2\sigma^2} -\frac{d}{2} \log(2\pi\sigma^2) \biggr].
\end{eqnarray*}
Note that the final term is a constant, given that $\sigma$ is assumed known.
Removing further terms independent of $A$ and $b$, we want to estimate
the transformation parameters that minimize
\[
\sum_{i=1}^{K+m} \sum_{j=1}^{K+n} p_{ji} \| x_j-A\mu_i-b \|^2 .
\]
Ignoring the terms independent of $b$, and noting that
$\partial a' x/\partial x = a$ and $\partial x' x/\partial x = 2x$,
the maximum likelihood estimates [Walker (\citeyear{Wal00})] are
%
%e9 ###
\begin{equation}
\hat{b}=\frac{\sum_{i=1}^{K+m} \sum_{j=1}^{K+n} p_{ji} (x_j-A\mu
_i)}{\sum_{i=1}^{K+m} \sum_{j=1}^{K+n} p_{ji}}
\label{eq2:bmle}
\end{equation}
and
%
%e10 ###
\begin{eqnarray}\label{eq2:Amle}
\hat{A}&=&\Biggl[ \sum_{i=1}^{K+m} \sum_{j=1}^{K+n} p_{ji} (x_j-\bar
{x})(\mu_i-\bar{\mu})' \Biggr]
\nonumber
\\[-8pt]
\\[-8pt]
\nonumber
&&{}\times
\Biggl[ \sum_{i=1}^{K+m} \sum_{j=1}^{K+n} p_{ji} (\mu_i-\bar{\mu})(\mu
_i-\bar{\mu})' \Biggr]^{-1}.
\end{eqnarray}

The algorithm alternates between the E-step and the M-step. At each
iteration, the transformation parameters are updated in the M-step to
$A^{(r+1)} = \hat{A}^{(r)}$ and $b^{(r+1)} = \hat{b}^{(r)}$,
before substitution into the E-step for the next iteration.

We assign convergence to be when $r$ is such that
%
%e11 ###
\begin{equation}
\frac{1}{(K+m+1)(K+n)}\sum_{i=0}^{K+m} \sum_{j=1}^{K+n} \bigl[
p_{ji}^{(r+1)}-p_{ji}^{(r)} \bigr]^2 \leq 10^{-l},
\label{eq2:Convergence}
\end{equation}
where $l$ is chosen and the posterior probability of $\mu_i$ matching
$x_j$ at the $r$th and $(r+1)$st iteration is denoted by $p_{ji}^{(r)}$
and $p_{ji}^{(r+1)}$, respectively, for $i=0,\ldots,K+m$ and $j=1,\ldots,K+n$.

%s3.2 ###
\subsection{Inference for $M$}\label{sec3.2}

Let $P$ be the $(K+n) \times(K+m+1)$ matrix containing the final
posterior matching probabilities. Let $\hat{A}$ and $\hat{b}$ be the
final~estimates of the transformation parameters obtained from the EM
algorithm.\looseness=-1

An obvious route to estimate the matching matrix, $M$, is to use the
posterior matching probabilities,
but this will not yield a one-to-one outcome.
For one-to-one matches we need to satisfy the constraints in (\ref
{eq2:MConstraint}) and (\ref{eq2:HardMConstraint}). Given the transformation,
the conditional log-likelihood of $M$ is
$\sum_{i=0}^{K+m} \sum_{j=1}^{K+n} M_{ij} \log P_{ji}$.
We find~$M$ that maximizes this log-likelihood by mixed integer linear
programming. In
our implementation
we imputted the $2K+m+n$ constraints into
lp\_solve [Berkelaar (\citeyear{Ber08})], which then yields the
estimated one-to-one matching matrix, $\hat{M}$.
We can summarize the steps as follows.

\begin{algo*}
\begin{enumerate}[(iii)]
\item[(i)]
Assign $q_{ij}$ using (\ref{eq2:qAdaptM}), (\ref{eq2:qAdapt0}) and (\ref
{eq2:qAdaptNM}) for $i=0,\ldots,K+m$ and $j=1,\ldots,K+n$.

\item[(ii)]
Find initial estimates of the transformation parameters, $A^{(0)}$ and
$b^{(0)}$, and assign the variance, $\sigma^2$. Possible choices are
discussed in the following subsection.

\item[(iii)]
Run the EM algorithm to get the updated estimates, $p_{ji}^{(1)}$,
$A^{(1)}$ and $b^{(1)}$, using~(\ref{eq2:Estep}), (\ref{eq2:Amle}) and
(\ref{eq2:bmle}), respectively.

\item[(iv)]
Repeat step 3 to find the updated estimates, $p_{ji}^{(r+1)}$,
$A^{(r+1)}$ and $b^{(r+1)}$, until convergence [defined in (\ref
{eq2:Convergence})] is reached. Let the final posterior matching
probabilities be stored in the $(K+n) \times(K+m+1)$ matrix~$P$ and
the final estimated transformation parameters be denoted by $\hat{A}$
and $\hat{b}$.

\item[(v)]
One-to-one matches are obtained using the hardening algorithm described above.

\item[(vi)]
Treating the matches within the inferred matching matrix, $\hat{M}$, as
known, we can update the transformation parameters using Procrustes
methodology [Dryden and Mardia (\citeyear{DryMar98})] to calculate the final estimates,
$\hspace*{3pt}\hat{\hspace*{-3pt}\hat{A}}$ and $\hspace*{1pt}\hat{\hspace*{-1pt}\hat{b}}$.
\end{enumerate}
\end{algo*}

%s3.3 ###
\subsection{Assigning the function and parameters within the EM algorithm}\label{sec3.3}

We need to assign the function $f$ stated in (\ref{eq2:qAdaptM}),
as well as starting values for the transformation parameters denoted by
$A^{(0)}$ and $b^{(0)}$,
and a variance $\sigma^2$.
We look at each assignment separately.

%s3.3.1 ###
\subsubsection{Distance function}\label{sec3.3.1}

As before, $\mu_j$ contains the allocated marker coordinates for marker~$j$ in $\mu$,
$j=1,\ldots,K$, and $\gamma_j$ is the index of the true marker~$j$ in
$\mu$.
Let $\bar{d}_{ij}$ denote the expected distance between a point $\mu_i$
and~$\mu_j$ for $i=1,\ldots,K+m$. Due to the freedom for a gel to warp,
in reality the distance between $\mu_i$ and $\mu_j$ in an image is
$d_{ij} = \bar{d}_{ij} + \varepsilon$,
where $\varepsilon$ denotes some error.

Our choice of the function, $f$, in (\ref{eq2:qAdaptM}), considers
all points in $\mu$ as possible true markers. We adopt a
multivariate normal distribution for $\varepsilon$, which gives
%
%e12 ###
\begin{equation}
q_{ij} = p(\gamma_j=i) \propto\exp\biggl\{ -\frac{\| \mu_i - \mu_j \|
^2}{2\sigma_*^2} \biggr\},
\label{eq2:qAdaptM2}
\end{equation}
for $i=1,\ldots,K+m$, where $\sigma_*^2$ is the variance between two
points in $\mu$ (assuming independence across dimensions). So the
probability that $\mu_i$ is the true marker $j$ will decrease the
further it is from $\mu_j$.

%s3.3.2 ###
\subsubsection{Starting values for transformation parameters}\label{sec3.3.2}

As we have prior knowledge of allocated corresponding markers in
both
$\mu$ and $x$, it is sensible that $A^{(0)}$\vadjust{\goodbreak} and $b^{(0)}$ are set as
the transformation parameters necessary to best superimpose
corresponding markers. Dryden and Mardia (\citeyear{DryMar98}) show how these
parameters can be estimated from the matrix,
%
%e13 ###
\begin{equation}
R = (\mu_*' \mu_*)^{-1} \mu_*' x^{(m)},
\label{eq2:A0b0}
\end{equation}
where $\mu_*$ is the $K \times(d+1)$ matrix $\mu_* = (\underline
{1}_K,\mu^{(m)})$ and $\underline{1}_K$ is a vector of ones of length
$K$. The $K \times d$ matrices, $\mu^{(m)}$ and $x^{(m)}$, contain only
the marker coordinates for $\mu$ and $x$, respectively.

The first column in $R'$ contains $b^{(0)}$ and the second two columns
in $R'$ contain the $d \times d$ matrix $A^{(0)}$.

%s3.3.3 ###
\subsubsection{Starting values for the variance between images}\label{sec3.3.3}

We can estimate the variance $\sigma^2$ by considering the mean squared
distance between corresponding markers in $\mu$ and $x$ after an affine
transformation has been applied to superimpose them. That is, set
%
%e14 ###
\begin{equation}
\hat\sigma^2 = \frac{1}{\nu} \sum_{j=1}^K \bigl\| x_j - A^{(0)}\mu_j
-b^{(0)} \bigr\|^2,
\label{eq2:sigma0}
\end{equation}
where $\nu= dK-d^2-d$ and denotes the degrees of freedom. Here $dK$ is
the number of error terms in the $d$ components of the $K$ markers.
This number is reduced in $\nu$ to accommodate the estimates of
$A^{(0)}$ and $b^{(0)}$.

%s4 ###
\section{Grossly misallocated or missing markers}\label{sec4}
This section describes further refinements to the above Composite Algorithm,
which is highly dependent on the transformation parameters input as
starting values,
$A^{(0)}$ and~$b^{(0)}$. We have previously stated that the affine
transformation necessary
to superimpose corresponding markers in $\mu$ and $x$ will provide sensible
starting values for the transformation parameters within the EM
algorithm. However,
this would not be the case if gross misallocations occur.
The number of missing or grossly misidentified markers are dependent on
the quality of the equipment and the expert that creates the images.

First, we provide a method that will highlight grossly misallocated
markers across images. Highlighted markers can then be automatically
removed or corrected before they are used within the EM algorithm to
estimate transformation starting values. Then, in Section \ref
{subsec:missing} we deal with the
case where some markers are missing from one of the images.

%s4.1 ###
\subsection{Grossly misallocated markers}\label{sec4.1}
Gross misallocations of a marker may occur through human error when
inputting marker labels into data spreadsheets. Dryden and Walker (\citeyear{DryWal99})
consider procedures based on S estimators, least median of squares and
least quartile difference estimators that are highly resistant to
outlier points. The RANSAC algorithm
[Fischler and Bolles (\citeyear{FisBol81})] uses a similar robust strategy. Here we
describe how we can use the EM algorithm previously described.\vadjust{\goodbreak}

Here we provide a method that will highlight grossly misallocated
markers across images. Highlighted markers can then be automatically
removed or corrected before they are used within the EM algorithm to
estimate transformation starting values.

Let $\mu^{(m)}$ and $x^{(m)}$ be $K \times d$ coordinate matrices where
$\mu_j$ and $x_j$ contain the coordinates of marker $j$ in $\mu$ and
$x$, respectively, for $j=1,\ldots,K$.
Here we consider the prior matching probabilities
to be independent of the distance between a possible marker and the
allocated marker so that
%
%e15 ###
\begin{equation}\label{eq2:qGross}
q_{ij}=
\cases{
p_M ,& \quad$\mbox{for } i=j,$ \vspace*{2pt}\cr
\displaystyle\frac{1-p_M}{K}, &\quad$\mbox{for } i \neq j,$}
\end{equation}
where $p_M$ denotes the probability that the allocated marker $\mu_j$
truly corresponds to the allocated marker $x_j$.

We input $\mu^{(m)}$ and $x^{(m)}$ into steps (i)--(v) of the composite
algorithm to estimate the one-to-one matching matrix $\hat{M}$,
replacing (\ref{eq2:qAdaptM}) and (\ref{eq2:qAdapt0}) with~(\ref{eq2:qGross})
in stage (i). We use (\ref{eq2:A0b0}) to estimate
the starting
transformation values, $A^{(0)}$ and $b^{(0)}$. Note that the starting
transformation will
be distorted by the presence of grossly misallocated markers.
There are four possible outcomes for $k=1,\ldots,K$:
\begin{itemize}
\item
The allocated corresponding markers $\mu_k$ and $x_k$ are matched if
$\hat{M}_{kk}=1$.
We include both $\mu_k$ and $x_k$ in further analyses.

\item
The marker $x_k$ remains unmatched if
$\hat{M}_{0k}=1$.
We exclude both $\mu_k$ and~$x_k$ from further analyses.

\item
No point in $x^{(m)}$ is matched to the marker $\mu_k$ if
$\hat{M}_{kj}=0$,
for all $j=1,\ldots,K$. We exclude both $\mu_k$ and $x_k$ from further analyses.

\item
The marker $\mu_{k_1}$ is matched to an allocated noncorresponding
marker $x_{k_2}$ if
$\hat{M}_{k_1 k_2} = 1$,
for $k_1 \neq k_2$. We exclude $\mu_{k_1}$, $\mu_{k_2}$, $x_{k_1}$ and
$x_{k_2}$ from further analyses.
\end{itemize}
See Section~\ref{subsec:eg1} for an illustration.
%s4.2 ###
\subsection{Missing markers}\label{subsec:missing}

It is possible that all $K$ markers are not successfully located in
both $\mu$ and $x$. For example, only 10 out of the possible $K=12$
markers were located in the image displayed in Figure~\ref{fig1:Image}.

There are four possible cases we must consider for Marker $k=1,\ldots
,K$: (a)~located in both $\mu$ and $x$; (b)
located in $\mu$ alone; (c)
located in $x$ alone; and (d)
not located in either $\mu$ or $x$.
We first introduce notation to allow for the possibility of missing markers.

Let $K_\mu$ and $K_x$ be the total number of markers located in $\mu$
and $x$, respectively. As previously noted, let $\mu$ be the $(K + m)
\times d$ coordinate matrix and $x$ be the $(K + n) \times d$
coordinate matrix.

If marker $k$ is located in $\mu$, then $\mu_k$ contains the
coordinates of marker~$k$ in $\mu$. If marker~$k$\vadjust{\goodbreak} is not located in $\mu
$, then $\mu_k = \varnothing$. Similarly, if marker~$k$ is located in
$x$, then $x_k$ contains the coordinates of marker~$k$ in $x$, for
$k=1,\ldots,K$. If marker $k$ is not located in~$x$, then $x_k =
\varnothing$.

As previously stated, $Q$ is the $(K+m+1) \times(K+n)$ matrix
containing the prior matching probabilities for points in $x$. We
define $Q$ by allowing for the possibility that an allocated marker $k$
is not the true marker $k$, for $k=1,\ldots,K$.

\textit{Markers in} $x$: corresponding to each of the above cases we
have:
\begin{longlist}[(a)]
\item[(a)] If $\mu_j \neq\varnothing$ and $x_j \neq\varnothing$, we
assign $q_{ij}$ as previously stated in (\ref{eq2:qAdaptM}) and~(\ref
{eq2:qAdapt0}) for $i=0,\ldots,K+m$.
\item[(b)] If $\mu_j \neq\varnothing$ and $x_j = \varnothing$, we treat
$\mu_j$ as a nonmarker.
\item[(c)] If $\mu_j = \varnothing$ and $x_j \neq\varnothing$, we treat
$x_j$ as a nonmarker.
\item[(d)] If $\mu_j = \varnothing$ and $x_j = \varnothing$, we set
$q_{ij}=q_{jk}=\varnothing$ for all $i$ and $k$.
\end{longlist}

\textit{Nonmarkers in} $x$:
The prior matching probability of a nonmarker, $x_j$, is again set to
be uniform over all matching possibilities so that, for $i=0,\ldots
,K+m$ and $j=K+1,\ldots,K+n$,
%
%e16 ###
\begin{equation}
q_{ij} = \frac{1}{K_\mu+m+1}.
\label{eq2:qAdaptNM2}
\end{equation}
In case 3, when $\mu_j = \varnothing$ and $x_j \neq\varnothing$ for
$j=1,\ldots,K$, we treat $x_j$ as a nonmarker and use (\ref
{eq2:qAdaptNM2}) to calculate $q_{ij}$ for $i=0,\ldots,K+m$.

Note that $\mu$ contains $K_\mu$ markers and $m$ nonmarkers. There are
only $K_\mu+m+1$ matching possibilities for a point in $x$, thus
producing the denominator in (\ref{eq2:qAdaptNM2}).
See Section~\ref{subsec:eg2} for an illustration.

%s5 ###
\section{Examples}\label{sec5}
Our full data set---see \hyperref[sec:supp]{Supplementary Material}
[Mardia, Petty and Taylor (\citeyear{MPT12})]---was
collected to represent eight subjects, under two different conditions,
treated with two possible treatments.

A replicate image was also produced for each subject-treatment specific case.
A~typical Western Blot is shown in Figure~\ref{fig1:Image}, which is
approximately of
size $280 \times220$. In this paper we illustrate the methods on two
pairs of images:
in the first example, robustness to gross misidentification is
explored, and the second
example deals with missing markers.
%s5.1 ###
\subsection{Grossly misallocated marker}\label{subsec:eg1}

Let $\mu$ and $x$ represent the coordinate sets on Western Blots of
a renal cancer cell line cultured under either normoxic or hypoxic
conditions. The proteins are then extracted and probed with either
patient sera or control sera in a Western Blot to produce the images generated.
%a control treated with normoxia and
%a renal cancer patient treated with normoxia, respectively.
All $K=12$ markers were located in both images.

We input the corresponding markers for $\mu$ and $x$ into steps
(i)--(v) of the
composite algorithm (see
Section~\ref{sec3.2}) to estimate the one-to-one matching matrix, $\hat{M}$, found
when superimposing $\mu^{(m)}$ onto $x^{(m)}$. That is, we transform
the appropriate markers in $\mu$ onto the corresponding markers in $x$.
Using only the markers,\vadjust{\goodbreak} we estimate the variance in~(\ref{eq2:x|M}) as
$\hat{\sigma}^2 = 4.5^2$ and set
the prior matching probability in~(\ref{eq2:qGross}) as $p_M=0.99$.
The starting values for the transformation parameters, $A^{(0)}$ and
$b^{(0)}$, are found using (\ref{eq2:A0b0}). We use the final
posterior probabilities, $P$, to estimate $M$.
Marker $1$ remains unmatched in both images.
%
%f3 ###
\begin{figure}

\includegraphics{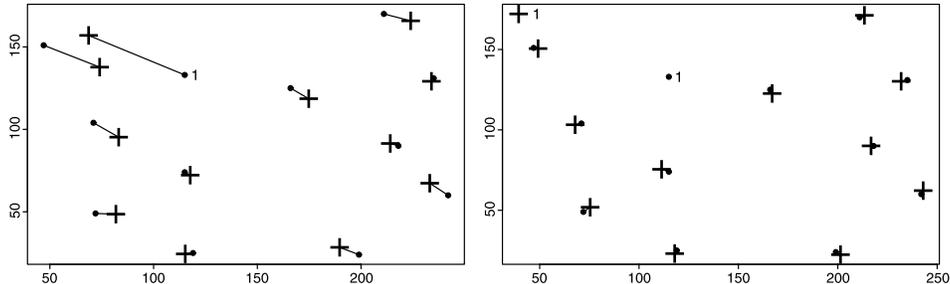}

\caption{Initial transformation, before (left) and after (right)
marker $1$ is removed as a~marker from both images.}\label{fig3:GrossReal1}
\end{figure}

Figure~\ref{fig3:GrossReal1} shows the initial transformation of $\mu$
onto $x$ before and after marker~$1$ is removed as a marker (though
still displayed) in both images. In this example, the RMSD between the
$12$ marker pairs before the removal is $19.44$. The RMSD between the
remaining $11$ marker pairs after the removal is $2.96$.

Following these discoveries, we were told that marker $1$ was
incorrectly labeled as spotID 136 when it should have been spotID 153,
that is, the methodology was able to highlight a misidentified marker.

%s5.2 ###
\subsection{Missing markers}\label{subsec:eg2}

In this example we display the matches made when comparing two
\textit{replicate} specimens,
representing a
cell line cultured under either normoxic conditions, with proteins
extracted and probed with control sera. All $12$ markers were located
in $\mu$. Markers $9$ and $10$ were missing in $x$, so these were
treated as
nonmarkers in $\mu$ and we set $K=10$.

\begin{figure}

\includegraphics{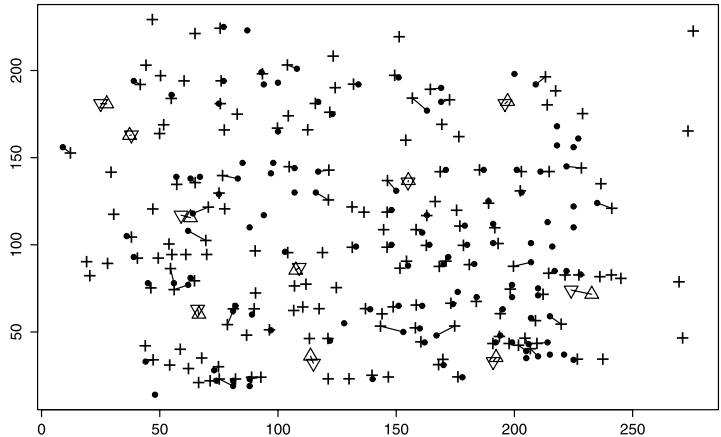}

\caption{Final transformation of $\mu$ onto $x$ and the matches made.
Points in $x$ ($\bullet$), points in transformed $\mu$ ($+$), markers
in $x$ ($\triangledown$) and
markers in $\mu$ ($\triangle$). The 107 matched points across images
are joined by a line.}\label{fig3:MatchEx}
\end{figure}

We input the images into steps (i)--(v) of the composite algorithm. The
starting values for the transformation parameters, $A^{(0)}$ and
$b^{(0)}$, are found using (\ref{eq2:A0b0}). We estimate the variance
in (\ref{eq2:x|M}), $\sigma^2$, using (\ref{eq2:sigma0}) with
denominator~$\nu$. Finally, we set $\hat{\sigma}_*^2 = \hat{\sigma}^2$
in (\ref{eq2:qAdaptM2}).
The estimated transformation parameters are
\[
\hat{A} =
\pmatrix{
0.980 & -0.047 \vspace*{2pt}\cr
0.002 & 1.006}
\]
and $\hat{b}=(-1.72,10.78)'$.
We display the matches made in Figure~\ref{fig3:MatchEx} after the
final transformation of $\mu$ onto $x$.
%
%f4 ###

%s6 ###
\section{Discussion}\label{sec6}

Many EM algorithms are known to converge only to a~local solution, and
this will also apply
to the methods considered here. However, the availability of
the\vadjust{\goodbreak}
markers which provide
partial information will usually ensure good starting values, so this
will not be a
problem in our application.

Note that it would be possible to adapt the model so that
$\sigma$ could be allowed to vary according to the distance of the
point to
the edge of the image, which could be used to deal with minor nonlinear
deformations.
More generally, it should also be possible to adapt our methods to deal
with more
general transformations, for example, using thin-plate splines [Chui and
Rangarajan (\citeyear{ChuRan03})].

There are situations when clusters occur within a gel which makes it
difficult to correctly identify a marker within a cluster of points. We
can allow for the increased likelihood that a marker $\mu_j, j=1,\ldots
,K$, is misallocated if it
exists within a cluster of other points, by using an adaptive choice of
$f$ in the
prior (\ref{eq2:qAdaptM}):
\[
q_{ij}=p(\gamma_j=i) \propto
\cases{
\displaystyle\frac{1}{C_j}, &\quad $\mbox{if } d_{ij} \leq\varepsilon,$\vspace*{2pt}\cr
0, &\quad $\mbox{if } d_{ij} > \varepsilon,$}
\]
where $d_{ij}$ is the Euclidean distance and
$C_j$ is the number of points in $\mu$ that are within a distance of
$\varepsilon$ from $\mu_j$, that is,
\[
C_j = \sum_{i=1}^{K+m} I[d_{ij} \leq\varepsilon],
\]
where $I[d_{ij} \leq\varepsilon]=1$ if $d_{ij} \leq\varepsilon$, ($0$
otherwise) for $i=1,\ldots,K+m$.

A further adaptation of the model, which could be useful in Western Blots,
would be to incorporate in the priors a measure associated with the\vadjust{\goodbreak}
grey-scale intensity of the located points in the image [Rohr, Cathier and W{\"o}rz (\citeyear{RohCatWor04})].
Approaches for this, as well as further models for the
background noise, are
considered in Petty (\citeyear{Pet09}).

Our composite algorithm ensures one-to-one matches, but there are
circumstances in
which many-to-one or many-to-many matches
can be considered. These can be useful when comparing protein images
since multiple forms of an individual protein can often be visualized
[Banks et al. (\citeyear{Banetal00})]. That is, a single protein can produce
multiple spots on an image.

It should be noted that our model is asymmetric in $\mu$ and $x$. This
is not uncommon;
for example, the full Procrustes error is not symmetrical [see Dryden
and Mardia (\citeyear{DryMar98})]. Also, the standard RMSD used by bioinformatricians
is again not a symmetrical measure. However, there are symmetrical
unlabeled shape analyses; see Green and Mardia (\citeyear{GreMar06}), for example.
However, this method has not been developed for affine transformations
and warping as
required here. There is also a nonprobabilistic method of Rangarajan, Chui and Bookstein (\citeyear{RanChuBoo97}) for
similarity shape, but again the extension of the method to affine
transformations
and warping requires further work; see Kent, Mardia and Taylor (\citeyear{KenMarTay10N2}) for a
statistical
framework. For the data considered here, we have verified that
reversing the role of $\mu$ and $x$ does not change the broad conclusions.

Finally, we note that the methods described in this paper could have
applications in other situations in which there are unlabeled points,
some of which---possibly with error---have been manually identified.
Thus, the method could be used in the preparation of ground truth for training
an object recognition system or a pose estimation system; for example,
see the survey of
Murphy-Chutorian and Trivedi (\citeyear{MurTri08}).

\section*{Acknowledgments}
We would like to thank Roz Banks and Rachel Craven for providing us
with gel data and general discussion concerning protein gels. We would
also like to thank David Hogg for useful
references about further applications.

%%%%%%%%%%%%%%%%%%%%%%%%%%%%%%%%%%%%%%%%%%%%%%
%

\begin{supplement}[id=suppA]
\label{sec:supp}
\stitle{Western Blot data}
\slink[url]{http://lib.stat.cmu.edu/aoas/544/blots.tar.gz}
\slink[doi]{10.1214/12-AOAS544SUPP}
\sdatatype{.gz}
\sdescription{The supplementary data contains a zipped file which
includes information taken from 28
Western Blots. This represents 8 subjects (four controls and four
patients) treated with two
possible treatments. A replicate image is also obtained for each
subject-treatment combination,
though some replicates are missing. Further details are included in the
associated README file.}
\end{supplement}

%suskaldyti doi
% imsref loaded by akundreckaite, 2012-04-11 12:49:03

\printaddresses

\end{document}